\documentclass{amsart}
\usepackage{graphicx}
\usepackage{epsf}

\usepackage{amsmath, amssymb,latexsym,float,algorithm,algorithmic}
\usepackage{enumerate}

\theoremstyle{definition}

\theoremstyle{remark}

\numberwithin{equation}{section}
\newcommand{\norm}[1]{\left\Vert#1\right\Vert}

\newcommand{\Real}{\mathbb R}
\newcommand{\Complex}{\mathbb C}

\newcommand{\R}{\text{\fontshape{n}\selectfont I\kern-.42exR}}

\newcommand{\1}{\text{\fontshape{n}\selectfont 1\kern-.56exl}}

\begin{document}

\centerline{\Large\it UKQCD Collaboration and University of Edinburgh}

\title[Computational Methods for UV-Suppressed Fermions]
{Computational Methods for UV-Suppressed Fermions}

\author{Artan Bori\c{c}i}



\date{20 Aug 2002}
\maketitle
\begin{abstract}
Lattice fermions with suppressed high momentum modes solve
the ultraviolet slowing down problem in lattice QCD.
This paper describes a stochastic evaluation of the
effective action of such fermions. The method is a based
on the Lanczos algorithm and it is shown to have the same
complexity as in the case of standard fermions.
\end{abstract}
\pagebreak

\section{Introduction}

There has been recent interest in the so-called `Ultraviolet slowing
down' of fermionic simulations in lattice QCD
\cite{Irving_et_al,Duncan_et_al,PhdF,Peardon,AHasenfratz_Knechtli}.
These studies try to address algortmically large fluctuations
of the high end modes of the fermion determinant. The goal is to
increase the signal-to-noise ratio of the infrared modes and to
accelerate fermion simulations as well.

In fact, all the computational effort needed to treat UV-modes
by above algorithms can be reduced to zero by suppressing them
in the first place \cite{Borici_UVSF}. The lattice Dirac
operator of this fermion theory is given by:
\begin{equation}\label{D_operator}
D = \frac{\mu}{a} \Gamma_5 \tanh \frac{a \Gamma_5 D_{W/S/O}}{\mu}
\end{equation}
where $D_{W/s/o}$ is the input lattice Dirac operator, $a$ the
lattice spacing and $\mu > 0$ is a dimensionless parameter.
For Wilson (W) and overlap (o) fermions as the input theory one has
$\Gamma_5 = \gamma_5$. For staggered fermions $\Gamma_5$ is a diagonal
matrix with entries $+1/-1$ on even/odd lattice sites.
The theory converges to the input theory in the contimuum limit
and is local and unitary as shown in detail in
\cite{Borici_UVSF}. The input theory is also recovered in the limit
$\mu \rightarrow \infty$. For $\mu \rightarrow 0$ one has
$D \rightarrow \mu$, i.e. a quenched theory.

Perturbative calculations with this theory are straightforward.
To fix the idea I assume in the following
Wilson fermions to be input fermions.
The inverse fermion propagator is given by
\begin{equation}\label{inverse_propagator}
{\tilde D}(p) = \frac{\mu}{a} \gamma_5 \tanh \frac{a {\tilde
H_W}(p)}{\mu}
\end{equation}
with $p = \{p_{\nu}, \nu=1,\ldots,4\}$ being the four-momentum
vector.
As usual, gauge fields are parametrized by $su(3)$
elements:
\begin{equation}\label{su3}
U(x)_{\nu} = e^{i a g A(x)_{\nu}}, ~~~~A(x)_{\nu} \in su(3)
\end{equation}
and the Wilson operator is written as a sum of
the free and interaction terms:
\begin{equation*}
D_W = D_W^0 + D_W^I
\end{equation*}
The splitting of the lattice Dirac operator is written
in the same form:
\begin{equation*}
D = D^0 + D^I,
~~~~D^0 = \frac{\mu}{a} \gamma_5 \tanh \frac{a H_W^0}{\mu}
\end{equation*}
where the interaction term has to be determined.
This can be done by expanding $D$ in terms of $a/\mu$:
\begin{equation}\label{pert_expan}
D = D_W[\1 + c_1 (\frac{a H_W}{\mu})^2
           + c_2 (\frac{a H_W}{\mu})^4 + \cdots]
\end{equation}
where $c_1, c_2, \ldots$ are real expansion coefficients.
Calculation of $D^I$ is an easy task if one stays with a finite
number of terms in the right hand side of (\ref{pert_expan}).
Also, the number of terms can be minimized using a
Chebyshev approximation for the hyperbolic tangent.
\footnote{I would like to thank Joachim Hein for discussions
related to lattice perturbation theory.}

In this paper I describe computational methods needed to
evaluate the effective action of the theory defined above.
In particular, the complexity of the proposed Lanczos method
does not depend on the input sparse matrix that describes a
fermion theory on the lattice.

In the following section
I derive a class of Lanczos based methods for computations
with the proposed theory and then in section 3 conclusions
follow.

\section{Lanczos based methods for computations with fermions}

The effective action of the theory defined above can be written as:
\begin{equation}\label{eff_action}
S_{\text{eff}} = \text{tr} f(A)
\end{equation}
where $A \in \Complex^{N\times N}$ and $f(s)$
is a real and smooth function of $s \in \Real^{+}$. The matrix $A$
is assumed to be Hermitian and positive definite.
Since the trace is difficult to obtain one can use the stochastic
method of \cite{Bai_et_al}. The method is based on evaluations of
many bilinear forms of the type:
\begin{equation}\label{b_forms}
{\mathcal F}(b,A) = b^T f(A) b
\end{equation}
where $b \in \Real^N$
is a random vector. The trace is estimated as an average over many
bilinear forms. A confidence interval can be computed as described
in detail in \cite{Bai_et_al}.

The method described here is similar to the method
of \cite{Bai_et_al}. Its viability for lattice QCD computations
has been demonstrated in the recent work of \cite{Cahill_et_al}.
\cite{Bai_et_al} derive their method using quadrature rules and Lanczos
polynomials. Here, I give an alternative derivation which
uses familiar tools (in lattice simulations) such as
sparse matrix invertions and Pad\'e approximations.
The Lanczos method enters the derivation as an algorithm
for solving linear systems of the form:
\begin{equation}\label{lin_sys}
A x = b, ~~~~~~x \in \Complex^N
\end{equation}

\subsection{Lanczos algorithm}

I follow standard standard texts as \cite{Golub_VanLoan} and
notations and arguments of \cite{Borici_over,Borici_WUP,Borici_isqr}.
$n$ steps of the Lanczos algorithm \cite{Lanczos} on the
pair $(A,b)$ are given by Algorithm \ref{Lanczos_algor}.
\begin{algorithm}[htp]
\caption{The Lanczos algorithm}
\label{Lanczos_algor}
\begin{algorithmic}
\STATE Set $\beta_0 = 0, ~q_0 = o, ~q_1 = b / ||b||^2$
\FOR{$~i = 1, \ldots n$}
    \STATE $v = A q_i$
    \STATE $\alpha_i = q_i^{\dag} v$
    \STATE $v := v - q_i \alpha_i - q_{i-1} \beta_{i-1}$
    \STATE $\beta_i = ||v||_2$
    \STATE $q_{i+1} = v / \beta_i$
\ENDFOR
\end{algorithmic}
\end{algorithm}

The Lanczos vectors $q_1, \ldots, q_n \in \Complex^N$
can be compactly denoted by the matrix $Q_n = [q_1, \ldots, q_n]$.
They are a basis of the Krylov subspace
$\mathcal{K}_n = \text{span}\{b,Ab,\ldots,A^{n-1}b\}$.
It can be shown that the following identity holds:
\begin{equation}\label{AQ_QT}
A Q_n = Q_n T_n + \beta_n q_{n+1} e_n^T, ~~~~~q_1 = b/||b||_2
\end{equation}
$e_n$ is the last column of the identity matrix ${\1}_n \in \Real^{n\times n}$
and $T_n$ is the tridiagonal and symmetric Lanczos matrix (\ref{T_n})
given by:
\begin{equation}\label{T_n}
T_n =
\begin{pmatrix} \alpha_1 & \beta_1  &             &             \\
                \beta_1  & \alpha_2 & \ddots      &             \\
                         & \ddots   & \ddots      & \beta_{n-1} \\
                         &          & \beta_{n-1} & \alpha_n    \\
\end{pmatrix}
\end{equation}
The matrix (\ref{T_n}) is usually referred to as the Lanczos matrix.
Its eigenvalues, the so called Ritz values, tend to approximate the
extreme eigenvalues of the original matrix $A$.

To solve the linear system (\ref{lin_sys}) I seek an approximate solution
$x_n \in \mathcal{K}_n$ as a linear combination of the
Lanczos vectors:
\begin{equation}\label{x_n}
x_n = Q_n y_n, ~~~~~y_n \in \Complex^n
\end{equation}
and project the linear system (\ref{lin_sys}) on to the
Krylov subspace $\mathcal{K}_n$:
\begin{equation*}
Q_n^{\dag} A Q_n y_n = Q_n^{\dag} b = Q_n^{\dag} q_1 ||b||_2
\end{equation*}
Using (\ref{AQ_QT}) and the orthonormality of Lanczos vectors, I obtain:
\begin{equation*}
T_n y_n = e_1 ||b||_2
\end{equation*}
where $e_1$ is the first column of the identity matrix ${\1}_n$.
By substituting $y_n$ into (\ref{x_n}) one obtains the approximate
solution:
\begin{equation}\label{x_sol}
x_n = Q_n T_n^{-1} e_1 ||b||_2
\end{equation}

\subsection{Algorithms for the bilinear form (\ref{b_forms})}

The theoretical framework of the algorithm of \cite{Bai_et_al}
can be based on
the Pad\'e approximant of the smooth and bounded
function $f(.)$ in an interval. The Pad\'e approximation
can be expressed as a partial fraction expansion. Therefore,
one can write:
\begin{equation}
f(s) \approx \sum_{k=1}^m \frac{c_k}{s + d_k}
\end{equation}
with $c_k \in \Real, d_k \geq 0, k = 1, \ldots, m$.
It is assumed that the right hand side converges to the left hand side
when the number of partial fractions becomes large enough.
For the bilinear form I obtain:
\begin{equation}\label{partial_frac}
{\mathcal F}(b,A) \approx \sum_{k=1}^m b^T \frac{c_k}{A + d_k \1} b
\end{equation}

A first algorithm can already be written down at this point.
Having computed the
partial fraction coefficients one can use a multi-shift iterative
solver of \cite{Freund} to evaluate the right hand side (\ref{partial_frac}).
To see how this works, I solve the shifted linear system:
\begin{equation*}
(A + d_k \1) x^k = b
\end{equation*}
using the same Krylov subspace $\mathcal{K}_n$. A closer inspection
of the Lanczos algorithm, Algorithm \ref{Lanczos_algor} suggests that
in the presence of the shift $d_k$ I get:
\begin{equation*}
\alpha_i^k = \alpha_i + d_k
\end{equation*}
while the rest of the algorithm remains the same. This is the so called
shift-invariance of the Lanczos algorithm. From this property and by
repeating the same arguments which led to (\ref{x_sol}) I get:
\begin{equation}\label{xk_sol}
x^k_n = Q_n \frac{1}{T_n + d_k {\1}_n} e_1 ||b||_2
\end{equation}
Using the shift-invariance of the Lanczos algorithm
I obtain Algorithm \ref{shifted_sys}.

\begin{algorithm}[htp]
\caption{The Lanczos algorithm for solving $(A+d_k\1)x^k=b$.}
\label{shifted_sys}
\begin{algorithmic}
\STATE Set $\beta_0 = 0, ~\rho^1_1 = 1 / ||b||_2, ~q_0 = o, ~q_1 = \rho^1_1 b,
x^k_0 = o, ~{\tilde x}^k_0 = o, ~\rho^k_0 = 0$
\FOR{$~i = 1, \ldots$}
    \STATE $v = A q_i$
    \STATE $\alpha_i = q_i^{\dag} v$
    \STATE $v := v - q_i \alpha_i - q_{i-1} \beta_{i-1}$
    \STATE $\beta_i = ||v||_2$
    \STATE $q_{i+1} = v / \beta_i$
    \FOR{$~k = 1, \ldots, m$}
        \STATE ${\tilde x}^k_{i+1} = - ({\tilde x}^k_i \alpha_i
           + {\tilde x}^k_{i-1} \beta_{i-1}) / \beta_i$
        \STATE $\rho^k_{i+1} =
           - (\rho^k_i \alpha_i + \rho^k_{i-1} \beta_{i-1}) / \beta_i$
        \STATE $r^k_{i+1} = q_{i+1} / \rho^k_{i+1}$
        \STATE $x^k_{i+1} = {\tilde x}^k_{i+1} / \rho^k_{i+1}$
    \ENDFOR
    \IF{$1 / |\rho^1_{i+1}| < \epsilon$}
       \STATE $n = i$
       \STATE stop
    \ENDIF
\ENDFOR
\end{algorithmic}
\end{algorithm}
Note that
the residual errors $r^k_i, i=1,\ldots,n,k=1,\ldots,m$
are given by:
\begin{equation*}
r^k_i = b - A x^k_i - d_k x^k_i
\end{equation*}
In exact arithmetic their norm is given by:
\begin{equation}\label{rec_res}
1/\rho^k_i = \norm{b - A x^k_i - d_k x^k_i}_2
\end{equation}
By applying Algorithm \ref{shifted_sys}
one can solve the shifted linear systems on the right hand side of
(\ref{partial_frac}). The algorithm stops if the linear system
with the smallest shift is solved to the desired accuracy $\epsilon$.
This is a well-known technique \cite{Freund}
which is used also in lattice QCD \cite{Frommer_masses}.
However, the problem with
this method is that one needs to store a
large number of vectors that is proportional
to $m$. This could be prohibitive if $m$ is say larger than $10$.

In fact, the right hand side of (\ref{partial_frac}) can be written
in terms of solutions $x^k_n, k = 1, \ldots,m$ as a sum of scalars:
\begin{equation}
{\mathcal F}(b,A) \approx \sum_{k=1}^m c_k w^k, ~~~~~~~~~w^k = b^T x^k
\end{equation}
Therefore, it is easy to replace the vector recurrences
by scalar recurrences of the form:
\begin{equation}
{\tilde w}^k_{i+1} = - ({\tilde w}^k_i \alpha_i
 + {\tilde w}^k_{i-1} \beta_{i-1}) / \beta_i
\end{equation}

In this way one obtains the Algorithm \ref{shifted_sys2}.
It is clear that by applying Algorithm \ref{shifted_sys2}
one gains substantial storage savings
compared to Algorithm \ref{shifted_sys}.
If one has a good Pad\'e approximant for the function $f(.)$ one can
apply Algorithm \ref{shifted_sys2}.
\begin{algorithm}[htp]
\caption{The Lanczos algorithm for computing $w^k, k = 1, \ldots, m$.}
\label{shifted_sys2}
\begin{algorithmic}
\STATE Set $\beta_0 = 0, ~\rho^1_1 = 1 / ||b||_2, ~q_0 = o, ~q_1 = \rho^1_1 b,
w^k_0 = o, ~{\tilde w}^k_0 = o, ~\rho^k_0 = 0$
\FOR{$~i = 1, \ldots$}
    \STATE $v = A q_i$
    \STATE $\alpha_i = q_i^{\dag} v$
    \STATE $v := v - q_i \alpha_i - q_{i-1} \beta_{i-1}$
    \STATE $\beta_i = ||v||_2$
    \STATE $q_{i+1} = v / \beta_i$
    \FOR{$~k = 1, \ldots, m$}
        \STATE ${\tilde w}^k_{i+1} = - ({\tilde w}^k_i \alpha_i
           + {\tilde w}^k_{i-1} \beta_{i-1}) / \beta_i$
        \STATE $\rho^k_{i+1} =
           - (\rho^k_i \alpha_i + \rho^k_{i-1} \beta_{i-1}) / \beta_i$
        \STATE $w^k_{i+1} = {\tilde w}^k_{i+1} / \rho^k_{i+1}$
    \ENDFOR
    \IF{$1 / |\rho^1_{i+1}| < \epsilon$}
       \STATE $n = i$
       \STATE stop
    \ENDIF
\ENDFOR
\end{algorithmic}
\end{algorithm}
Note that one another way to save storage is
using the multi-shift Gonjugate Gradient variant of \cite{Beat_Jeg}.

\subsection{An exact method}

If a Pad\'e approximation is not sufficient or difficult to obtain,
the Lanczos method is the only alternative
to evaluate exactly the bilinear forms of type
(\ref{b_forms}).

To see how this is realized I assume that the linear system (\ref{lin_sys})
is solved to the desired accuracy using the Lanczos algorithm,
Algorithm {\ref{Lanczos_algor} and (\ref{x_sol}).
In the application considered here one can show that:
\begin{equation}\label{lemma}
\sum_{k=1}^m b^T \frac{c_k}{A + d_k \1} b = ||b||^2 \sum_{k=1}^m
e_1^T \frac{c_k}{T_n + d_k {\1}_n} e_1
\end{equation}
For the result (\ref{lemma}) to hold, it is sufficient to show that:
\begin{equation*}
b^T \frac{c_k}{A + d_k \1} b = ||b||^2
e_1^T \frac{c_k}{T_n + d_k {\1}_n} e_1
\end{equation*}
which can be shown using the
orthonormality property of the Lanczos vectors and (\ref{xk_sol}).
Note however that in presence of roundoff errors
the orthogonality of the Lanczos vectors is lost but the result
(\ref{lemma}) is still valid. The interested reader may
consult the work of \cite{Cahill_et_al,Golub_Strakos}.

From this result and the convergence of the partial fractions to the
matrix function $f(.)$, it is clear that:
\begin{equation}\label{reduced_form}
{\mathcal F}(b,A) \approx
{\mathcal {\hat F}}_n(b,A) = ||b||^2 e_1^T f(T_n) e_1
\end{equation}
Note that the evaluation of the right hand side
is a much easier task than
the evaluation of the right hand side of (\ref{b_forms}).
A straightforward method is the spectral decomposition of the
symmetric and tridiagonal matrix $T_n$:
\begin{equation}\label{Omega}
T_n = Z_n \Omega_n Z_n^T
\end{equation}
where $\Omega_n \in \Real^{n\times n}$ is a diagonal matrix of
eigenvalues $\omega_1,\ldots,\omega_n$ of $T_n$ and $Z_n \in \Real^{n\times n}$
is the corrsponding matrix of eigenvectors, i.e. $Z_n = [z_1,\ldots,z_n]$.
From (\ref{reduced_form}) and
(\ref{Omega}) it is easy to show that (see for example \cite{Golub_VanLoan}):
\begin{equation}\label{omega_form}
{\mathcal {\hat F}}_n(b,A) = ||b||^2 e_1^T Z_n f(\Omega_n) Z_n^T e_1
\end{equation}
where the function $f(.)$ is now evaluated at individual eigenvalues of
the tridiagonal matrix $T_n$.

The eigenvalues and eigenvectors of a symmetric and tridiagonal matrix
can be computed by the QL method with implicit shifts
\cite{Numerical_Recipes}. The method has an $O(n^3)$ complexity.
Fortunately, one can compute (\ref{omega_form}) with
only an $O(n^2)$ complexity.
Closer inspection of eq. (\ref{omega_form}) shows that besides the
eigenvalues, only the first elements of the eigenvectors are needed:
\begin{equation}\label{result_form}
{\mathcal {\hat F}}_n(b,A) = ||b||^2 \sum_{i=1}^n z_{1i}^2 f(\omega_i)
\end{equation}
It is easy to see that the QL method delivers the eigenvalues and
first elements of the eigenvectors with $O(n^2)$ complexity.
\footnote{I thank Alan Irving for the related comment on the QL implementation
in \cite{Numerical_Recipes}.}

A similar formula (\ref{result_form}) is suggested
by \cite{Bai_et_al}) based on
quadrature rules and Lanczos polynomials.
The Algorithm \ref{lambda_algor} is thus another way to compute the
bilinear forms of the type (\ref{b_forms}).
\begin{algorithm}[htp]
\caption{The Lanczos algorithm for computing (\ref{b_forms}).}
\label{lambda_algor}
\begin{algorithmic}
\STATE Set $\beta_0 = 0, ~\rho_1 = 1 / ||b||_2, ~q_0 = o, ~q_1 = \rho_1 b$
\FOR{$~i = 1, \ldots$}
    \STATE $v = A q_i$
    \STATE $\alpha_i = q_i^{\dag} v$
    \STATE $v := v - q_i \alpha_i - q_{i-1} \beta_{i-1}$
    \STATE $\beta_i = ||v||_2$
    \STATE $q_{i+1} = v / \beta_i$
    \STATE $\rho_{i+1} = - (\rho_i \alpha_i + \rho_{i-1} \beta_{i-1}) / \beta_i$
    \IF{$1 / |\rho_{i+1}| < \epsilon$}
       \STATE $n = i$
       \STATE stop
    \ENDIF
\ENDFOR
\STATE Set $~(T_n)_{i,i} = \alpha_i, ~(T_n)_{i+1,i} = (T_n)_{i,i+1} = \beta_i$,
       otherwise $~(T_n)_{i,j} = 0$
\STATE Compute $\omega_i$ and $z_{1i}$ by the QL method
\STATE Evaluate (\ref{b_forms}) using (\ref{result_form})
\end{algorithmic}
\end{algorithm}

Clearly, the Lanczos algorithm and Algorithm \ref{shifted_sys2}
has an $O(nN)$ complexity, whereas Algorithm \ref{lambda_algor} has
a greater complexity: $O(nN)+O(n^2)$. However, Algorithm \ref{lambda_algor}
delivers an exact evaluation of (\ref{b_forms}). For typical applications
in lattice QCD the $O(n/N)$ overhead is small and therefore
Algorithm \ref{lambda_algor} is the recommended algorithm among all three
algorithms presented in this section.

A remark on stopping criteria is also desirable. The method of
\cite{Bai_et_al} computes the relative differences of (\ref{result_form})
between two successive Lanczos steps and stops if they don't decrease
below a given accuracy. In order to perform the test their algorithm
needs to compute the eigenvalues of $T_i$ at each Lanczos step $i$.
This may be a large computational overhead.
On the other hand the test proposed here is theoretically
safe. This is clarified by the remark at the end of the proof
of the lemma (\ref{lemma}).
However, this test may be too prudent since the prime interest here
is the computation of the bilinear form (\ref{b_forms}).
\begin{figure}
\vspace{4cm}
\epsfxsize=6cm
\hspace{2cm} \epsffile[240 400 480 450]{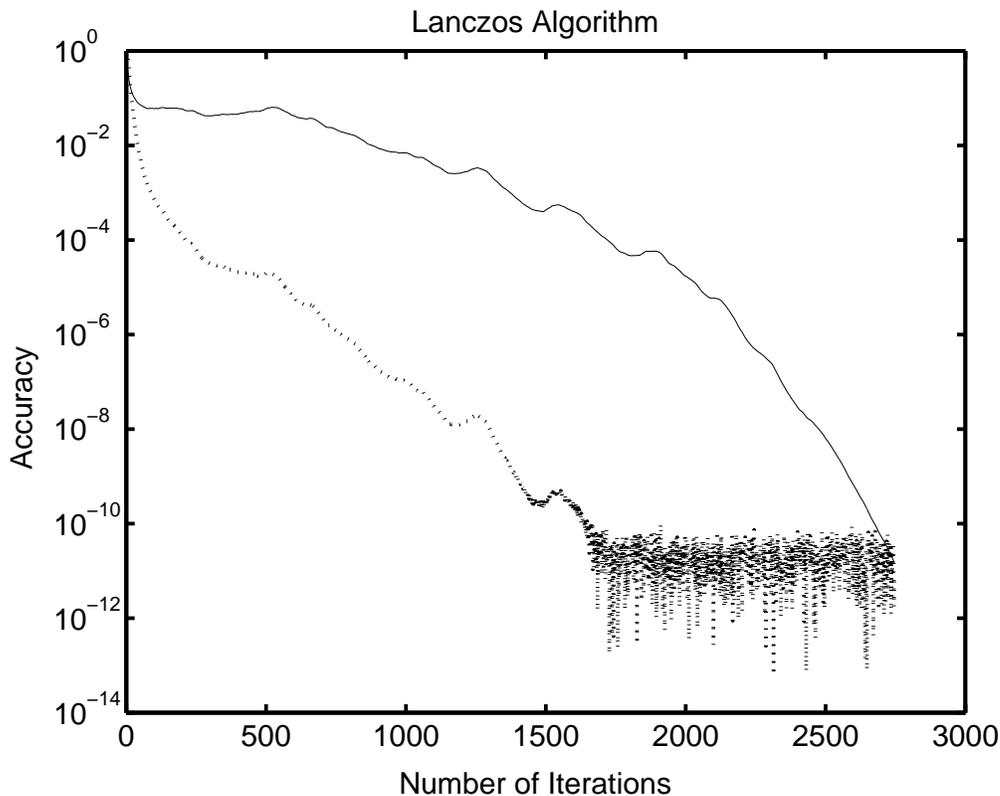}
\vspace{6cm}
\caption{Normalized recursive residual (solid line) and
relative differences of (\ref{result_form}) (dotted line)
produced by Algorithm \ref{lambda_algor}.}
\end{figure}

To illustrate this situation I give an example from a lattice
calculation (The lattice size and parameters are given in section 5.1).
I compute the bilinear form (\ref{b_forms}) for:
\begin{equation}
f(s) = \log \tanh \sqrt{s}, ~~~~~s \in \Real^{+}
\end{equation}
and $A = H_W^2$, $b \in \Complex^N$. The real and imaginary parts
of the $b-$elements are chosen randomly from the set $\{+1,-1\}$.

In Fig. 1 are shown the normalized recursive residuals
$\rho_0/\rho_i = \norm{b - Ax_i}_2/\norm{b}_2,i=1,\ldots,n$
and relative differences of (\ref{result_form})
between two successive Lanczos steps.
The figure illustrates clearly the different regimes of convergence
for the linear system and the bilinear form.
The relative differences of the bilinear form
converge faster than the computed recursive residual.
This example indicates that a stopping criterion based on the
solution of the linear system may indeed be strong.
Therefore, the recommended stopping criteria would be to monitor the
relative differences of the bilinear form
but less frequently than proposed by \cite{Bai_et_al}. More investigations
are needed to settle this issue.
Note also the roundoff effects (see Fig. 1) in the convergence of the
bilinear form.

\section{Conclusion}

In this paper I have described computational methods needed to
evaluate the effective action of the theory with suppressed
cutoff modes.

All methods described in this paper have a complexity that does
not depend on the input sparse matrix that describes the fermion
theory on the lattice. In this way, it may be concluded that
simulation algorithms of lattice QCD which are based on the
estimation of the effective fermion action have the same
complexity. The UV-suppressed fermions are such an example.

\section*{Acknowledgements}
I would like to thank Philippe de Forcrand and Alan Irving
for discussions and useful suggestions at different stages of this work.

\end{document}